\documentclass[%
 reprint,
superscriptaddress,
nofootinbib,
 amsmath,amssymb,
 aps,
 prc,
floatfix,
]{revtex4-2}
\usepackage{graphicx}
\usepackage{dcolumn}
\usepackage{bm}
\usepackage{hyperref}
\usepackage[mathlines]{lineno}

\usepackage{xcolor}
\usepackage{svg}
\usepackage{xspace}

\newcommand{\pp}[0]{$p$~+~$p$\xspace}

\newcommand{\pythia}[0]{\textsc{Pythia-8}\xspace}
\newcommand{\fastjet}[0]{\textsc{fastjet-3}\xspace}
\newcommand{\roounfold}[0]{\textsc{RooUnfold}\xspace}
\newcommand{\tenngen}[0]{\textsc{TennGen}\xspace}
\newcommand{\pT}{\ensuremath{p_{T}\xspace}\xspace}
\def\GeV{\ifmmode {\mathrm{\ Ge\kern -0.1em V}}\else
                   \textrm{Ge\kern -0.1em V}\fi \xspace}%
\newcommand{\akT}{anti-$k_{\mathrm{t}}$\xspace}

\newcommand{\vn}[1]{$v_{#1}$\xspace}
\newcommand{\psin}[1]{$\psi_{#1}$\xspace}

\newcommand{\jetvn}[1]{$v_{#1}^{\mathrm{jet}}$\xspace}
\newcommand{\jetpT}{$p_{\mathrm{T},\mathrm{jet}}$\xspace}

\newcommand{\FlowVec}[2]{\mathbf{#1}_{#2}}
\newcommand{\SingleEventAvgCorr}[2]{\langle #1_{#2} \rangle }
\newcommand{\EventAvgCorr}[2]{\langle \langle #1_{#2} \rangle \rangle }
\newcommand{\SingleEventAvgDiffCorr}[2]{\langle #1^{\prime}_{#2} \rangle }
\newcommand{\EventAvgDiffCorr}[2]{\langle \langle #1^{\prime}_{#2} \rangle \rangle }

\newcommand{\sNN}{$\sqrt{s_{{}_\mathrm{NN}}}$\xspace}

\begin{document}


\title{Reconstructing Jet Azimuthal Anisotropies with Cumulants} 

\author{Tanner Mengel} 
\affiliation{Department of Physics, University of Tennessee, Knoxville, TN, 37996, USA}
\affiliation{Department of Physics, University of Colorado, Boulder, Colorado 80309, USA}

\author{Niseem~Magdy} 
\affiliation{Department of Physics, University of Tennessee, Knoxville, TN, 37996, USA}
\affiliation{Department of Physics, Texas Southern University, Houston, TX 77004, USA}
\affiliation{Physics Department, Brookhaven National Laboratory, Upton, New York 11973, USA}

\author{Ron Belmont} 
\affiliation{Physics and Astronomy, University of North Carolina, Greensboro, North Carolina 27412, USA}

\author{Anthony Timmins} 
\affiliation{Department of Physics, University of Houston, Houston, Texas 77204, USA}

\author{Christine Nattrass} 
\affiliation{Department of Physics, University of Tennessee, Knoxville, TN, 37996, USA}

\date{\today}
\url{https://doi.org/10.1103/k68z-z2xr}

\begin{abstract}
In relativistic heavy-ion collisions, where quark-gluon plasma forms, hadron production is anisotropic at both low and high transverse momentum, driven by flow dynamics and spatial anisotropies. To better understand these mechanisms, we use multi-particle correlations to reconstruct jet anisotropies. We simulate data using \tenngen as a hydro-like background and combine it with \pythia generated jets, clustering them with the \akT algorithm. Jet anisotropies are unfolded using a Bayesian technique, ensuring the robustness of the reconstructed signals. Our results demonstrate that multi-particle cumulant methods can accurately capture the differential jet azimuthal anisotropies, providing crucial insights into high-\pT behavior and the dynamics within heavy-ion collisions.

\end{abstract}

\maketitle

\section{Introduction}

Relativistic heavy-ion collisions create an extremely hot and dense phase of matter known as the quark-gluon plasma (QGP). Investigations at the Relativistic Heavy Ion Collider (RHIC)~\cite{STAR:2005gfr, PHENIX:2004vcz, BRAHMS:2004adc, PHOBOS:2004zne}  and the Large Hadron Collider (LHC)~\cite{ALICE:2022wpn, CMS:2024krd,Busza:2018rrf} have demonstrated a wide array of complex phenomena related to the QGP. A key signature of QGP formation is the azimuthal anisotropy observed in the production of hadrons, particularly at low transverse momentum (\pT $<$ 3~\GeV{}). This anisotropy can be effectively modeled as hydrodynamic flow, with the final momentum distribution of hadrons reflecting initial spatial anisotropies, which are translated into momentum space through pressure gradients~\cite{Hirano:2005xf, Huovinen:2001cy, Hirano:2002ds, Romatschke:2007mq, Song:2010mg, Schenke:2011tv, Lacey:2013eia}. These azimuthal anisotropies are described using a Fourier expansion~\cite{Voloshin:1994mz, Poskanzer:1998yz}: 

\begin{eqnarray}
\frac{dN}{d\phi} &\propto&  C\left[ 1 + 2 \sum_{n=1}^{\infty} v_n \cos(n (\phi - \psi_n))\right]
\label{Eq:AzimuthalAnisotropy}
\end{eqnarray}

\noindent where $C$ is the normalization factor determined by the integral of the distribution, \vn{n} and \psin{n} denote the value and orientation for the anisotropy complex vector, respectively. In particular, \vn{2} and \vn{3} are called the elliptic and triangular coefficients, respectively.

At high transverse momentum (\pT $>$ 10~\GeV{}), hadron production in ion-ion (AA) collisions shows significant suppression compared to expectations based on proton-proton (\pp) collisions, a phenomenon understood as jet quenching~\cite{Connors:2017ptx, Qin:2015srf}. In this process, high-energy partons lose energy through both radiative and collisional interactions within the QGP~\cite{Qin:2015srf, Mehtar-Tani:2013pia, Blaizot:2015lma}. These high-\pT hadrons and their associated jets also exhibit non-zero azimuthal anisotropy, even though they fall outside the typical regime where the hydrodynamic flow would apply~\cite{CMS:2017xgk, ATLAS:2013ssy, ALICE:2018rtz}. Instead, these anisotropies are understood to arise from the spatial inhomogeneities in the QGP, with jet quenching effects being more pronounced for partons traveling through longer paths in the QGP~\cite{Gyulassy:2000gk, Shuryak:2001me, Jia:2012ez}. The pathlength is minimized when $\phi$ and $\psi_n$ are approximately aligned, leading to less jet attenuation.
In contrast, when $\phi$ and \psin{n} are approximately orthogonal, the pathlength is maximized, which leads to increased jet attenuation.

Both low- and high-\pT hadrons share a standard orientation of their anisotropies, correlating with the geometry of the colliding nucleons. A long-standing theoretical challenge has been to simultaneously explain both high-\pT suppression and the corresponding azimuthal anisotropy observed in A+A collisions, and several models have been proposed to address this issue~\cite{Shuryak:2001me, Molnar:2013eqa, Noronha-Hostler:2016eow, Betz:2016ayq, Holtermann:2024vdw, Barreto:2022ulg, Zhang:2013oca}. Previous measurements of jet differential azimuthal anisotropies \jetvn{n} have used an event plane method where the event plane is defined by soft particles ~\cite{ATLAS:2013ssy,ALICE:2015efi, Luong:2021hfl}. Given observations in the soft sector, these jet anisotropy coefficients should also be sensitive to the measurement technique.~Some studies indicate that the difference could be up to 10\%~\cite{Andres:2019eus,Zigic:2022xks,Zigic:2021rku}, which is on the order of the uncertainty on measurements by ATLAS~\cite{ATLAS:2021ktw}.

Measurements from smaller collision systems, such as \pp and $p$+Pb at the LHC~\cite{ATLAS:2012cix, CMS:2010ifv, CMS:2015yux, ALICE:2012eyl, CMS:2025kzg,  ATLAS:2019vcm}, and $p$+Au, $d$+Au, and ${}^{3}$He+Au at RHIC~\cite{PHENIX:2018lia, STAR:2022pfn}, have also revealed significant azimuthal anisotropies at low-\pT, with patterns similar to those observed in heavy-ion collisions. These findings suggest that even smaller systems may produce short-lived droplets of the QGP; indeed, hydrodynamic models have successfully described the low-\pT behavior in such systems. However, when it comes to high-\pT behavior in smaller collision systems, the expected signatures of jet quenching have not been observed. 
While some measurements indicate that no jet suppression at high transverse momentum in $p$+Pb and $d$+Au collisions~\cite{PHENIX:2015fgy, PHENIX:2020alr, ATLAS:2022iyq, ALICE:2017svf}, other studies indicate there may be some suppression~\cite{PHENIX:2023dxl}. No alterations in the momentum balance of dijets or hadron-jet pairs have been detected.

The ATLAS experiment has also reported non-zero azimuthal anisotropies for hadrons with \pT up to 12~\GeV{}, suggesting an extension of anisotropy effects into the high-\pT regime typically associated with jet quenching. It is unclear if differential jet quenching contributes to the observed high-\pT anisotropies~\cite{ATLAS:2022iyq}. This leaves two related unresolved puzzles: the lack of jet quenching observed in $p$+A collisions and the mechanism that produces high-\pT hadron anisotropies in the absence of jet quenching.

Further investigation into the mechanism responsible for high-\pT hadron anisotropies requires two key approaches: (i) comparing data models with both theory and experiment using consistent techniques and (ii) expanding measurements to incorporate multi-particle correlations. 
Discrepancies between event plane and cumulant methods can reach up to 10\%~\cite{Andres:2019eus, Zigic:2022xks, Zigic:2021rku, STAR:2021mii}. These variations can complicate the interpretation of experimental measurements made with the event plane method. 
Several factors, such as long-range non-flow effects and initial/final state fluctuations, can influence the flow correlations measured with the $2$-particle correlation method, potentially amplifying the observed flow signal. Therefore, applying $2$- and multi-particle cumulant methods in data analysis and theoretical calculations is crucial for a deeper understanding of the mechanisms behind high-\pT hadron anisotropies.

In this work, we demonstrate the feasibility of measuring the jet azimuthal anisotropies \jetvn{n} with $2$- and $4$-particle cumulants in Au+Au collisions simulated with \pythia~\cite{Sjostrand:2006za, Sjostrand:2007gs} jets merged with a \tenngen~\cite{Hughes:2020lmo, Mengel:2023mnw, TennGen} background.

\section{Method}

\subsection{Simulation}\label{sec:model}

Events are simulated in \tenngen~\cite{Hughes:2020lmo, TennGen, TennGen200}, a data-driven background generator designed to reproduce correlations arising from flow. In this model, the $n^{\text{th}}$-order event plane is set to zero for even values of \(n\), while for odd values of \(n\), the event plane is chosen randomly. \tenngen multiplicities and charged particle ratios match experimentally measured values~\cite{Aamodt:1313050, Abelev:2013vea}. Similar constraints are also applied to the transverse momentum spectra of charged pions, kaons, and protons through random sampling from an initial distribution fitted to a Blast Wave~\cite{Ristea:2013ara, Schnedermann:1993ws} distribution. In \tenngen, the generated particle $v^{}_n$($\eta$) is sampled from a flat distribution, an acceptable approximation for the region \( |\eta| < 1.1 \). The single particle $v^{}_{n}(p_{T})$ are modeled using a polynomial fit to the measured $v^{}_n(p_{T})$~\cite{PHENIX:2014uik, Adam:2016nfo}. \tenngen events used in this study are tuned to Au+Au collision data at \sNN = 200~\GeV{} for 20--30\% central events. \tenngen serves as a hydro-like background for jet interactions. The jet signal is generated using \pythia~\cite{Sjostrand:2007gs} with the Monash Tune 13~\cite{Skands:2014pea} and a minimum hard scattering transverse momentum cutoff of $\hat{p}_{T}^{\text{min}} > 5$ \GeV{}.

\begin{figure}[!htbp]
    \centering
    \includegraphics[width=0.94\columnwidth]{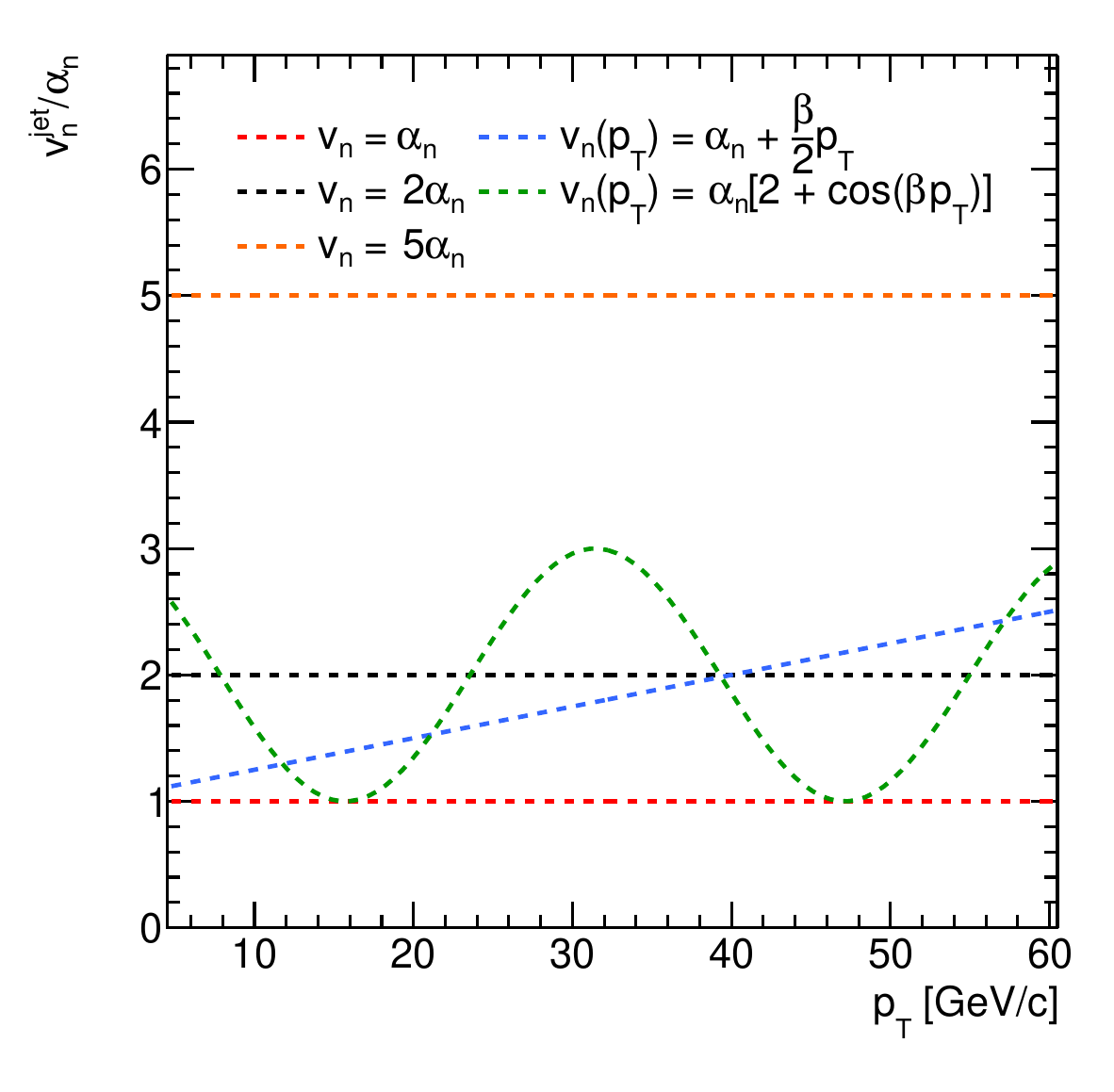}
    \caption{The dependence of \jetvn{n} on jet $p_{T}$ for functions used to realign leading PYTHIA jets in \pythia+\tenngen events. These include three $p_T$ independent constant values of $v^{\text{jet}}_{n}$ (red, black, and orange dotted lines), a $v^{\text{jet}}_{n}(p_T)$ which increases with $p_T$ (blue dotted line), and a physically unrealistic sinusoidal dependence on $p_T$ (green dotted line). The $y$-axis is scaled by $a_n$. }
    \label{fig:vnfunctions}
\end{figure}

The leading PYTHIA jets are realigned to simulate a $\frac{dN_{\text{jet}}}{d\phi_{\text{jet}}}$ distribution with non-zero differential azimuthal anisotropies \jetvn{n}. Previous measurements of $v^{\text{jet}}_{n}$ by the ATLAS collaboration indicate there may be a nonzero dependence on jet $p_{T}$ ~\cite{ALICE:2015efi,ATLAS:2013ssy,ATLAS:2021ktw}. To ensure our method is robust to such dependence, we consider several functional forms to model a wide range of momentum dependence on the jet azimuthal anisotropies \jetvn{n}$(p_{T})$. These include three $p_T$ independent constant values of $v^{\text{jet}}_{n}$, a $v^{\text{jet}}_{n}(p_T)$ which increases with $p_T$, and a physically unrealistic sinusoidal dependence on $p_T$. We are confident that any expected dependence of $v^{\text{jet}}_{n}$ on jet $p_{T}$ in experimental data will be less extreme than the latter. Leading PYTHIA jets are realigned to simulate nonzero \jetvn{2}, \jetvn{3}, and \jetvn{4} according to one of the following descriptions:
\begin{eqnarray}\label{eq:1-1}
    v^{\text{jet}}_{n} &=& \alpha_{n}, \nonumber \\
    v^{\text{jet}}_{n} &=& 2\alpha_{n},\nonumber \\
    v^{\text{jet}}_{n} &=& 5\alpha_{n},\nonumber \\
    v^{\text{jet}}_{n} &=& \alpha_{n} + \frac{\beta}{2}p_{T}, \nonumber\\
    v^{\text{jet}}_{n} &=& \alpha_{n}\left[2+\cos{\beta p_{T}} \right], 
\end{eqnarray}
\noindent where $\alpha_n$ is a nominal scale for each harmonic $n$, and $\beta$ is constant parameter for all $n$. The values used in this study are $a_2 = 4\%, a_3 = 1\%$, $a_4 = 0.5\%$, and $\beta = 0.2$. Figure~\ref{fig:vnfunctions} shows the five $v^{\text{jet}}_{n}$ used in the realignment of leading PYTHIA jets. A total of $10$ million \tenngen+\pythia events are generated for each sample. The simulation package developed to generate these events is publicly available~\cite{JetVnCumulant}.

The merged \pythia jet constituents and \tenngen charged particles are clustered into \akT jets with jet resolution parameter $R=0.4$ using the energy recombination scheme implemented with \fastjet~\cite{Cacciari:2011ma}. The charged final-state hadrons from \pythia p+p events are clustered separately before merging with \tenngen, which defines the truth jet momentum. The \pp event is used if there is at least one jet with \jetpT$>$ 10~\GeV{}.  Ghost particles with $p_{T}^{\text{ghost}} = 1$ MeV are used to estimate the area of clustered merged jets as described in the FastJet manual~\cite{Cacciari:2011ma}.

The momentum of the merged jets is corrected using the jet-multiplicity background subtraction method~\cite{Mengel:2023mnw}
\begin{equation}
p_{T,jet}^{\text{corr, N}} = p_{T,jet}^{\text{raw}} - \rho_{\text{mult}}(N_{\text{tot}}- \langle N_{\text{sig}} \rangle), \label{Eq:pTN}
\end{equation}
where $N_{\text{tot}}$ is the observed number of particles within the jet, $\langle N_{\text{Sig}} \rangle$ is approximated by the average number of \pythia particles in a \tenngen+\pythia jet of a given matched uncorrected jet momentum bin, and $\rho_{\text{Mult}}$ is the mean transverse momentum per background particle.
Merged jet candidates satisfying $p_{T,jet}^{\text{corr}} > 5$ \GeV{} and $A_{\text{jet}} > 0.6 \pi R^{2}$~\cite{Steffanic:2023cyx} are geometrically matched back to the jet axis of the \pythia signal jet. The closest merged jet is considered a match if $\Delta R < 0.75R$ and the corresponding \pythia jet momentum $p_{T,jet}^{\text{PYTHIA}}$ is taken to be the truth momentum.

\subsection{Cumulant Method}
The traditional cumulant method~\cite{Bilandzic:2010jr, Jia:2017hbm}, is used to construct the $v^{}_{n}\{2k\}(p_{T})$, where $k$ is a positive integer. In this work, the $v^{\text{jet}}_{n}\{2k\}$ is given as:
\begin{eqnarray}\label{eq:2-1}
    v^{\text{jet}}_{n}\{2\} &=& \dfrac{\langle\langle 2^{\prime}_{n} \rangle\rangle }{ \langle\langle 2_{n} \rangle \rangle^{1/2}}, \nonumber \\  
   \langle\langle 2^{\prime}_{n} \rangle\rangle  &=& \langle  \langle e^{\textit{i} n (\psi^{A}_{1} -  \varphi^{C}_{2} )} \rangle \rangle \nonumber, \\ 
   \langle\langle 2_{n} \rangle\rangle  &=& \langle  \langle e^{\textit{i} n (\varphi^{C}_{1} -  \varphi^{C}_{2} )} \rangle \rangle, 
\end{eqnarray}
\begin{eqnarray}\label{eq:2-2}
   v^{\text{jet}}_{n}\{4\} &=& \dfrac{\langle \langle 4^{\prime}_{n} \rangle \rangle }{  \left( -\langle\langle 4_{n} \rangle \rangle \right)^{3/4} }, \nonumber \\ 
   \langle\langle 4^{\prime}_{n} \rangle \rangle  &=& \langle  \langle e^{\textit{i} n (\psi^{A}_{1} + \varphi^{C}_{2} - \varphi^{C}_{3} - \varphi^{C}_{4})} \rangle \rangle \nonumber \\ 
   &-& 2~\langle  \langle e^{\textit{i} n (\varphi^{C}_{1} -  \varphi^{C}_{2} )} \rangle \rangle    \langle  \langle e^{\textit{i} n (\psi^{A}_{1} (p_{T}) -  \varphi^{C}_{2} )} \rangle \rangle, \nonumber \\ 
   \langle\langle 4_{n} \rangle \rangle  &=& \langle  \langle e^{\textit{i} n (\varphi^{C}_{1} + \varphi^{C}_{2} - \varphi^{C}_{3} - \varphi^{C}_{4})} \rangle \rangle \nonumber \\ 
   &-& 2~\langle  \langle e^{\textit{i} n (\varphi^{C}_{1} -  \varphi^{C}_{2} )} \rangle \rangle    \langle  \langle e^{\textit{i} n (\varphi^{C}_{1} -  \varphi^{C}_{2} )} \rangle \rangle,
\end{eqnarray}
where $\varphi^{C}$ is the azimuthal angle of particles in the region $C$, $\psi^{A}$ is the azimuthal angle of jets in the region $A$, and  $\langle\langle 2_{n}\rangle\rangle$ is the event averaged $2$-particle correlations of order $n$. The prime (${}^{\prime}$) symbol is used to differentiate the particle of interest correlation from the reference correlation. The integrated $\varphi$ is taken from particles labeled as reference flow particles in region $C$ with $3.1<|\eta_{C}|<5.1$. The $\psi$ is the particle of interest (i.e., jet) azimuthal angle defined within region $A$ with $|\eta_{\text{A}}| < 1.1 - R$. More details about the traditional cumulant method can be found in Appendix~\ref{App1}.


The differential azimuthal anisotropies for jets  $v^{\text{jet}}_{n}\{2k\}$ are binned as a function of uncorrected jet momentum. Jet measurements with $p_{T}$ dependencies are subject to bin-by-bin migration from smearing due to finite resolution in the uncorrected measurement. This smearing can skew the event-averaged differential particle correlations if the functional form of $v^{\text{jet}}_{n}$ is dependent on jet momentum. If the $v^{\text{jet}}_{n}$ are independent of jet momenta, then this smearing will not change the results. Measurements of $v^{\text{jet}}_{n}$ indicate there may be a nonzero dependence on \jetpT, meaning that measurements using a cumulant method will need to be unfolded~\cite{ALICE:2015efi,ATLAS:2013ssy,ATLAS:2021ktw}.

We unfold the reconstructed differential single-event $n^{\mathrm{th}}$-order $2k$-particle correlations using the Bayesian unfolding method~\cite{DAGOSTINI1995487} in \roounfold2.0.0~\cite{Brenner:2019lmf}. From Equations~\ref{eq:2-1}~\ref{eq:2-2}, we define the single-event $n^{\mathrm{th}}$-order $2k$-particle correlations, for use in constructing a response matrix. That is 
\begin{eqnarray}\label{eq:2-3}
   \langle 2_{n} \rangle  &=&  \langle e^{\textit{i} n (\varphi^{C}_{1} -  \varphi^{C}_{2} )} \rangle ,  \nonumber\\
   \langle 2^{\prime}_{n} \rangle  &=&  \langle e^{\textit{i} n (\psi^{A}_{1}-  \varphi^{C}_{2} )} \rangle, 
\end{eqnarray}
\begin{eqnarray}\label{eq:2-4}
   \langle 4_{n} \rangle   &=&   \langle e^{\textit{i} n (\varphi^{C}_{1} + \varphi^{C}_{2} - \varphi^{C}_{3} - \varphi^{C}_{4})} \rangle -2~ \langle 2_{n} \rangle^2, \nonumber \\
   \langle 4^{\prime}_{n} \rangle  &=&   \langle e^{\textit{i} n (\psi^{A}_{1} + \varphi^{C}_{2} - \varphi^{C}_{3} - \varphi^{C}_{4})} \rangle - 2~\langle 2_{n} \rangle\langle 2^{\prime}_{n} \rangle,  
\end{eqnarray}
where $\varphi^{C}$ is the azimuthal angle of particles in the region $C$, $\psi^{A}$ is the azimuthal angle of jets in the region $A$, binned in reconstructed jet $p_T$. Again, the prime (${}^{\prime}$) symbol is used to differentiate the jet azimuthal correlations from the reference particle correlations.

\pythia jets (truth jets) are matched to \pythia+\tenngen jets (reconstructed jets) and the momentum and azimuthal angle of the \pythia jet is taken as the truth momentum ($p_{T,jet}^{\text{truth}}$) and truth azimuthal angle ($\phi_{\text{jet}}^{\text{truth}}$). A reconstructed-level jet is considered a match to the closest truth-level jet in $\eta,\phi$ space if the separation is $dR < \frac{3}{4}R$. From the matched truth jet, the truth-level single event correlation $\SingleEventAvgDiffCorr{(2k)}{n}^{\text{truth}}$ is calculated. There are no matching criteria imposed between the reconstructed-level $\SingleEventAvgDiffCorr{(2k)}{n}^{\text{reco}}$ and the truth-level $\SingleEventAvgDiffCorr{(2k)}{n}^{\text{truth}}$. Both single-event correlations are calculated using the same $p_{T,jet}$ binning with a bin-width of 5\GeV{}.

We construct a four-dimensional response matrix, populated for bin migration of the yield as a function of the reconstructed-level $\SingleEventAvgDiffCorr{(2k)}{n}^{\text{reco}} \otimes p_{T,jet}^{reco}$ and the matched truth-level $\SingleEventAvgDiffCorr{(2k)}{n}^{\text{truth}} \otimes p_{T,jet}^{truth}$. We then unfold our reconstructed jet differential single-event $n^{\mathrm{th}}$-order $2k$-particle correlations. The unfolding procedure is repeated until the change in $\chi^2$ between the unfolded and truth $\SingleEventAvgDiffCorr{(2k)}{n}$ becomes less than the uncertainties of the measured $\SingleEventAvgDiffCorr{(2k)}{n}$. This typically occurs between 3-5 iterations, and more iterations beyond this threshold would yield diminishing increases in unfolded $\SingleEventAvgDiffCorr{(2k)}{n}$ resolution. The average value of the resulting unfolded single-event correlation is $\SingleEventAvgDiffCorr{(2k)}{n}^{\text{corr}}$ is calculated for each jet momentum bin and used to compute $v^{\text{jet}}_{n}(p_{T})$, according to Equations~\ref{eq:2-1}-~\ref{eq:2-2}. 

\begin{figure*}[!t]
    \centering
    \includegraphics[width=0.9\textwidth]{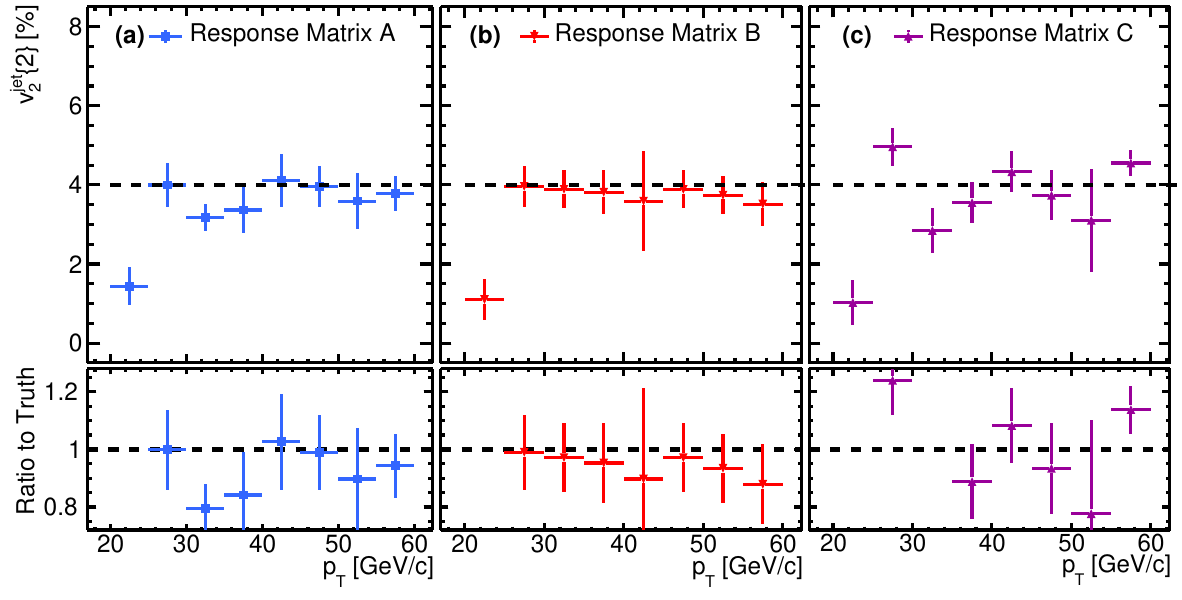}
    \caption{
    Comparisons of the $v^{\text{jet}}_{2}\{2\}$ calculated with unfolded $\SingleEventAvgDiffCorr{2}{2}^{\text{corr}}$ for  \pythia+\tenngen events with a $p_T$ independent truth $v_2^{\text{jet}} = 4\%$, unfolded with response matrices constructed from \pythia+\tenngen events where $v_2^{\text{jet}} = 4\%$ (Response Matrix A), a scaled constant $v_2^{\text{jet}} = 5\times 4\%$ (Response Matrix B), and a linear $p_T$ dependence $v_2^{\text{jet}}(p_{T}) = 4\% + (0.1\%)\times p_T$ (Response Matrix C). The dashed line represents the true constant $v^{\text{jet}}_{2} = 4\%$ of the unfolded sample.}
    \label{fig:unfolding_validation}
\end{figure*}

To demonstrate the robustness of this procedure, we unfold $\SingleEventAvgDiffCorr{(2k)}{n}^{\text{reco}}$ corresponding to a $p_T$ independent $v^{\text{jet}}_{n}$ with response matrices constructed from models with different functional forms. The results of this exercise are presented in Figure~\ref{fig:unfolding_validation}. We show that the corrected $v^{,\text{jet},\text{corr}}_{n}\{2k\}(p_{T})$ recovers the functional form of a given model, regardless of the functional form present in the construction of the response matrix. 

\section{Results}\label{sec:results}

\begin{figure}[!htbp]
    \centering
    \includegraphics[width=0.94\columnwidth]{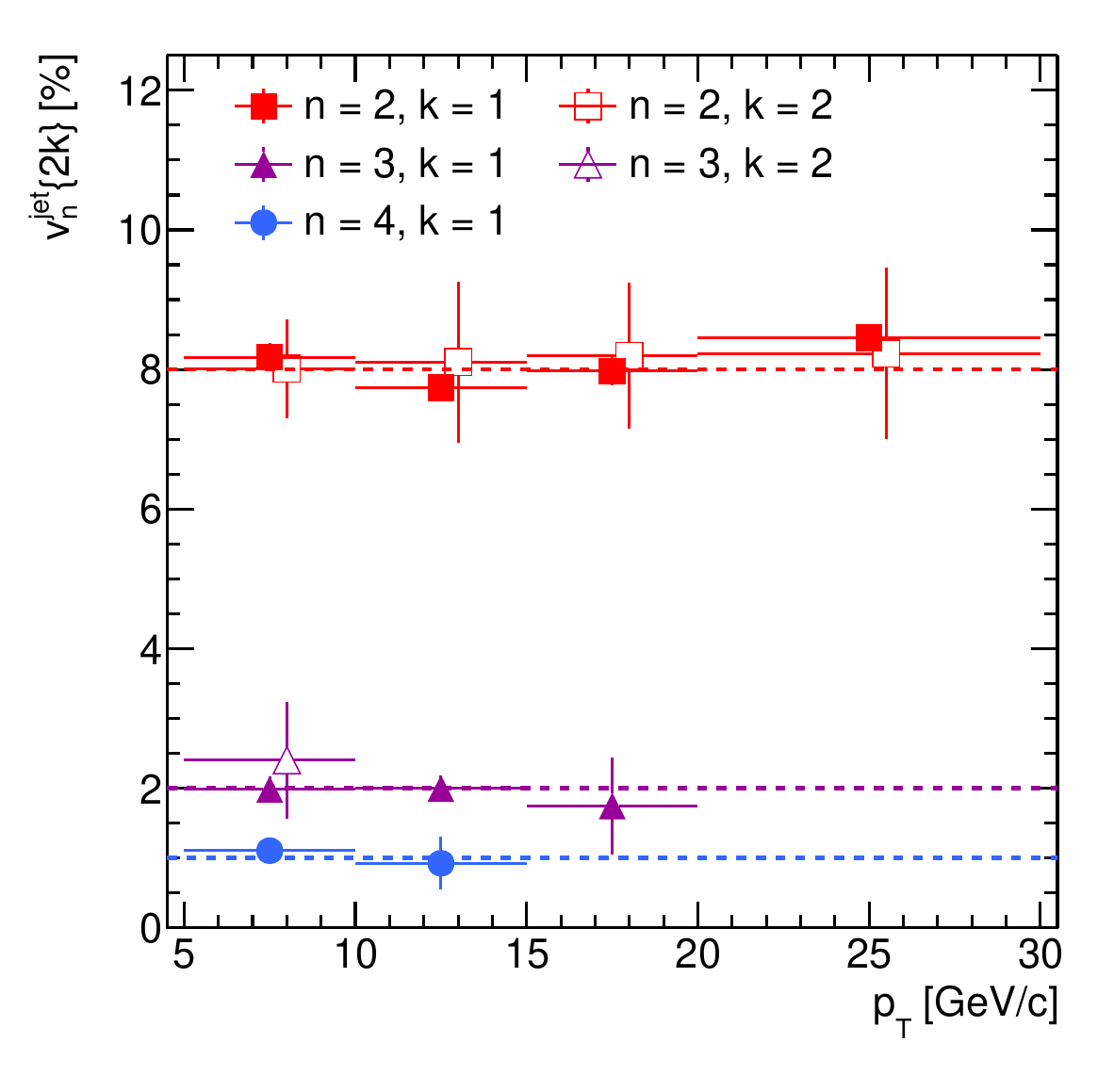}
    \caption{The \pT dependence of $v^{\text{jet}}_{n}\{2k\}$ for harmonic order $n = 2, 3$, and $4$, represented by the red squares, purple triangles, and blue circles, respectively. Values calculated using $2$- and $4$-particle correlations are differentiated by full ($k=1$) or open ($k=2$) symbols. The dashed line represents the true $v^{\text{jet}}_{n}(p_{T})$ values.}
    \label{fig:vn_three_panel}
\end{figure}

The simulated data using \pythia+\tenngen are analyzed using the $2$- and $4$-particle cumulant methods with several input flow signals. We present the $v^{\text{jet}}_{n}\{2\}$ for several $p_{T,jet}$ selection for Au+Au at 200 \GeV{} in Figure~\ref{fig:vn_three_panel}. The input $v^{\text{jet},\text{MC}}_{n,\text{jet}}$ is a constant equal to $0.08$, $0.02$, and $0.01$ for $n =$ 2, 3 and 4, respectively. The $v^{\text{jet},\text{corr}}_{n}\{2k\}$ are obtained after five iterations of Bayesian unfolding. Our results reflect the capabilities of using the $2$-particle cumulant method to reflect the input jet $v^{\text{jet}}_{n}$. The presented results also reflected the input sensitivity to the flow harmonic order attenuation.

Extending our analysis to the $4$-particle cumulant will offer deeper insights into the impact of flow fluctuations on jet flow measurements, as well as the effects of short- and long-range correlations. In this context, short-range correlations primarily arise from localized effects, such as particle decays or intra-jet correlations~\cite{Lacey:2005qq, Borghini:2000cm, Luzum:2010fb, Retinskaya:2012ky, ATLAS:2012at, Zhu:2005qa}. These contributions can be mitigated by applying multi-particle sub-event cumulant methods.

\begin{figure}[!htbp]
    \includegraphics[width=0.94\columnwidth]{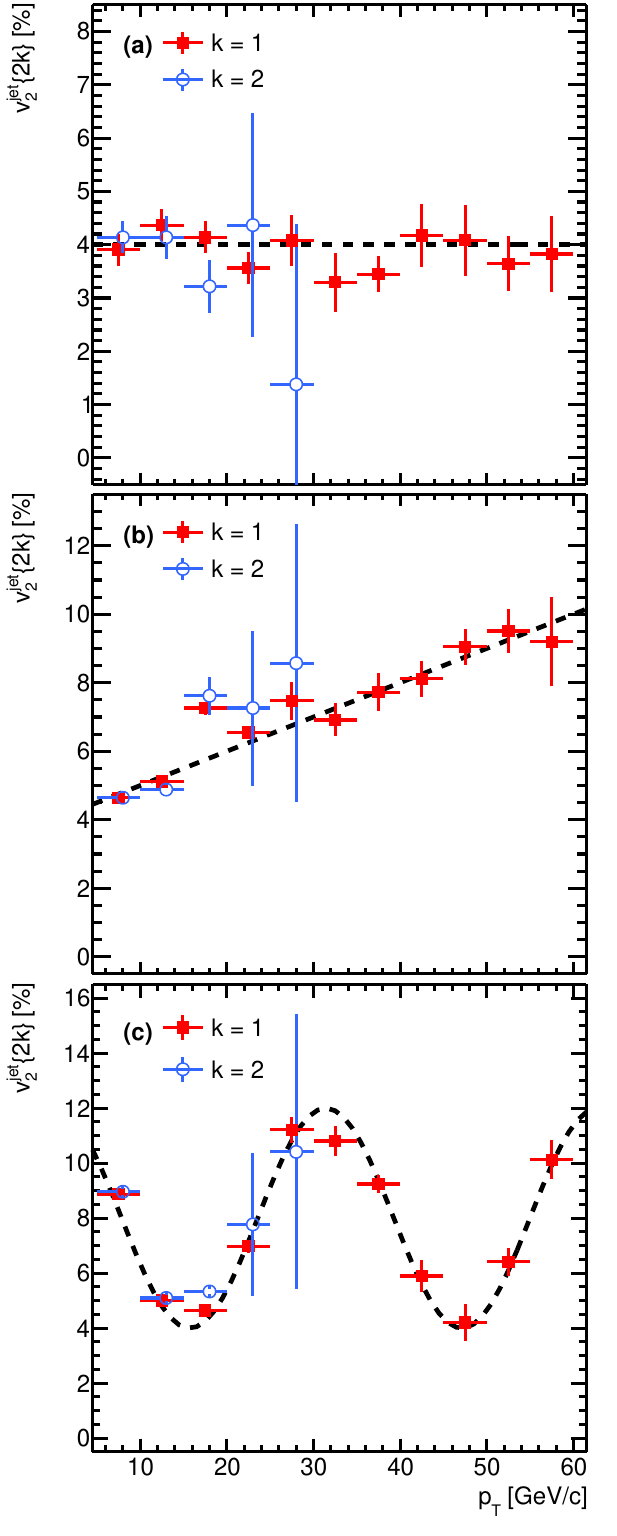}
    \caption{Comparison of the \pT dependence of the $v^{\text{jet}}_{2}\{2k\}$ for Au+Au collisions at 200~\GeV{} calculated with $2$-particle ($k=1$) and $4$-particle ($k=2$) correlations. The dashed line represents the true $v^{\text{jet}}_{n}(p_{T})$ values. The unfolded $v^{\text{jet}}_{2}\{2k\}$ $v_2^{\text{jet}} = 4\% $ $\textbf{(a)}$, a linear $p_T$ dependence $v_2^{\text{jet}}(p_{T}) = 4\% + (0.1\%)\times p_T$ $\textbf{(b)}$, and a sinusoidal $p_T$ dependence $v^{\text{jet}}_{n} = 4\%\times[2+\cos{0.2 p_{T}}]$ $\textbf{(c)}$.}
    \label{fig:v2_six_panel}
\end{figure}

Figure~\ref{fig:v2_six_panel} presents the corrected $v^{\text{jet}}_{2}\{2k\}$ (where $k=$ 1 or 2 for the 2- and 4-particle correlation, respectively) as a function of $p_{T,jet}$ for various functional forms of $v^{\text{jet},\text{MC}}_{n}$. These forms encompass different realistic scenarios where $v^{\text{jet}}_{n}$ remains either independent of or gradually increases with $p_{T,jet}$, as well as physically unrealistic forms, such as $v^{\text{jet}}_{n} \propto \cos(p_{T,jet})$. The inclusion of both realistic and unrealistic models underscores the robustness of our method in reconstructing differential jet azimuthal anisotropies, even under extreme conditions. Furthermore, our results indicate an agreement between $v^{\text{jet}}_{2}\{2\}$ and $v^{\text{jet}}_{2}\{4\}$, reflecting the model's nature, which lacks flow fluctuations effects.

\section{Conclusions}\label{sec:conclusions}

This work demonstrates the utility of $2$- and $4$-particle cumulant methods for measuring jet anisotropies, leveraging TennGen and PYTHIA models to simulate data representative jet backgrounds in the quark-gluon plasma. Our findings demonstrate that cumulants can be effectively used to reconstruct jet azimuthal anisotropies, capturing their flow characteristics and providing a robust means to analyze high-\pT jet azimuthal anisotropies. This methodology offers promise for disentangling long and short-range correlations between jets.  Since it is not possible to reconstruct an event plane in small systems this technique can be used to advance the understanding of jet interactions and anisotropies in heavy-ion and smaller collision systems. 

In real heavy-ion collisions, there are several effects, such as uncorrelated jets and resonance decays. These effects could be significant in small systems, and would have to be considered in order to properly interpret the results.  Additionally, event-by-event flow fluctuations will contribute to differences in different order cumulants. In A+A systems, where the relative fluctuations are mostly small, $v_{2}\{2\} = \sqrt{v_{2}^{2}+\sigma^{2}}$ and $v_{2}\{4\} \approx v_{2}\{6\} \approx v_{2}\{8\} \approx \sqrt{v_{2}^{2}-\sigma^{2}}$. However, in small systems, the relative fluctuations can be large, and the nature of the underlying distribution becomes important. Addressing these effects will be important for ensuring the applicability of this method to experimental data.

\section{Acknowledgments}
This work was supported in part by funding from the Division of Nuclear Physics of the U.S. Department of Energy under Grant No. DE-FG02-96ER40982. This work was performed on the computational resources at the Infrastructure for Scientific Applications and Advanced Computing (ISAAC) supported by the University of Tennessee.

\clearpage
\appendix
\section{Additional Cumulant Method Details}\label{App1}

We use the method presented in~\cite{Bilandzic:2010jr} which calculates multi-particle cumulants in terms of moments of $\FlowVec{Q}{}$ vectors defined as
\begin{equation}
    \FlowVec{Q}{n} \equiv \sum_{k=1}^{M} e^{in\phi_k}, \label{Eq:Qvec}
\end{equation}
where $M$ is the multiplicity of the referenced sub-event and $\phi_k$ is the azimuthal angle of particles labeled as Reference Flow Particles (RFP). We define our sub-event region as being away from mid-rapidity, $3.1 < |\eta| < 5.1$. For this study, we simulate events with $M$ fixed at 363. This $M$ corresponds to an average 20-30$\%$ central Au+Au $\sqrt{s_{NN}} = 200 $ \GeV{} collision~\cite{Aamodt:1313050}. We use a single bracket to denote average moments of Equation~\ref{Eq:Qvec} $\SingleEventAvgCorr{2k}{n}$, which are averaged over all $M$ and double brackets to denote event-averaged moments $\EventAvgCorr{(2k)}{n}$ which are averaged over all events. Using moments of Equation~\ref{Eq:Qvec}, the single event $n^{\mathrm{th}}$-order  $2$- and $4$-particle reference azimuthal correlations can be expressed as 
\begin{equation}
    \SingleEventAvgCorr{2}{n}  = \frac{|\FlowVec{Q}{n}|^2 - M}{M(M-1)},
\label{Eq:single_event_avg_corr_2part}
\end{equation}
and
\begin{equation}
 \begin{aligned}
    \SingleEventAvgCorr{4}{n}
     = & \frac{
    |\FlowVec{Q}{n}|^2 
    + |\FlowVec{Q}{2n}|^2 
    - 2\Re \{\FlowVec{Q}{2n} \FlowVec{Q}{n}^* \FlowVec{Q}{n}^* \}
    }{M(M-1)(M-2)(M-3)} \\
    & - 2\frac{
    (M-2)|\FlowVec{Q}{n}|^2 - M(M-3)
    }{M(M-1)(M-2)(M-3)},
\end{aligned}
\label{single_event_avg_corr_4part_Qvec}
\end{equation}

\noindent where $|\FlowVec{Q}{n}|^2$ is the squared magnitude of $\FlowVec{Q}{}$, $\FlowVec{Q}{2n}$ is the $\FlowVec{Q}{}$-vector for $n= 2n$, $\Re\{\cdot\}$ is the real component of the complex argument and $M$ is the reference multiplicity. The event averaged $n^{\mathrm{th}}$-order $2k$-particle reference azimuthal correlation $\EventAvgCorr{(2k)}{n}$ is computed with a weighted sum of the single event reference azimuthal correlation
\begin{equation}
 \begin{aligned}
    \EventAvgCorr{(2k)}{n} = \frac{1}{W\{2k\}}\sum_{i=1}^{N_{\text{events}}} (w\{2k\}){}_i (\SingleEventAvgCorr{(2k)}{n})_i,
\end{aligned}
\label{Eq:event_avged_corr}
\end{equation}
where $(w\{2k\})_{i}$ are event weights and $W\{2k\}$ is the sum of all event weights. Common choices for weights are reference multiplicity or particle momentum. In the case when the multiplicity of the reference is limited, methods for momentum weighting are outlined in~\cite{Bilandzic:2010jr}.
In our model, $M$ is fixed, and the $v_n$ of TennGen is independent of multiplicity. Therefore we use the weights 
\begin{equation}
\begin{aligned}
    &w\{2\} = M(M-1),\\
    &w\{4\} = M(M-1)(M-2)(M-3),\\
\end{aligned}
\label{Eq:event_avged_corr_weights}
\end{equation}
for $2$- and $4$-particle RFP correlations respectively.

The second-order cumulant for the reference sub-event is defined as
\begin{equation}
\begin{aligned}
    c_{n}\{2\} \equiv \EventAvgCorr{2}{n},
\end{aligned}
\label{Eq:cn2_def}
\end{equation}
\noindent which is simply the event averaged $n^{\mathrm{th}}$-order $2$-particle reference azimuthal correlation defined in Equation~\ref{Eq:single_event_avg_corr_2part}. The fourth-order reference cumulant is given by 
\begin{equation}
\begin{aligned}
    c_{n}\{4\} \equiv \EventAvgCorr{4}{n} - 2\cdot \EventAvgCorr{2}{n}^2, 
\end{aligned}
\label{Eq:cn4_def}
\end{equation}
which is the  $n^{\mathrm{th}}$-order $4$-particle reference azimuthal correlation corrected for event-by-event fluctuations in the  $n^{\mathrm{th}}$-order $2$-particle reference azimuthal correlation.

The differential single event azimuthal correlation for jets with respect to the reference correlation is calculated using  additional flow vector $\FlowVec{p}{}$, as well as the reference flow vector $\FlowVec{Q}{}$ defined in Equation~\ref{Eq:Qvec}. To measure the jet $v_n$, the Point of Interest (POI) vector $\FlowVec{p}{}$ is built using the azimuthal angle of all jets in an event within $|\eta_{\text{jet}}| < 1.1 - R$, defined as
\begin{equation}
    \FlowVec{p}{n} \equiv \sum_{k=1}^{m_p} e^{in\psi_k} ,\label{Eq:pvec}
\end{equation}
where $m_p$ is the number of jets within a given \pT bin in an event and $\psi_k$ is the azimuthal angle of the jet axis. 

The $n^{\mathrm{th}}$-order $2$- and $4$-particle differential single event azimuthal correlations are expressed as  
\begin{equation}
 \begin{aligned}
    \SingleEventAvgDiffCorr{2}{n} = \frac{\FlowVec{p}{n}\FlowVec{Q}{n}^*}{m_pM},
\end{aligned}
\label{Eq:single_event_avg_diff_corr_2part_def}
\end{equation}
and
\begin{equation}
 \begin{aligned}
    \SingleEventAvgDiffCorr{4}{n}  & =
    \frac{(|\FlowVec{Q}{n}|^2-2M+2)\FlowVec{p}{n}\FlowVec{Q}{n}^{*} -\FlowVec{p}{n}\FlowVec{Q}{n}\FlowVec{Q}{2n}^{*}}{m_pM(M-1)(M-2)},
\end{aligned}
\label{Eq:single_event_avg_diff_corr_4part_def}
\end{equation}
where $m_p$ is the number of jets within a given \pT bin in an event, and $\FlowVec{p}{n}$ is the POI vector. The event averaged  $n^{\mathrm{th}}$-order $2k$-particle reference differential azimuthal correlation $\EventAvgDiffCorr{(2k)}{n}$ is computed similar to Equation~\ref{Eq:event_avged_corr} with the corresponding weights 
\begin{equation}
\begin{aligned}
    &w^{\prime}\{2\} = m_pM,\\
    &w^{\prime}\{4\} = m_pM(M-1)(M-2),\\
\end{aligned}
\label{Eq:event_avged_diff_corr_weights}
\end{equation}
for the differential $2$- and $4$-particle azimuthal correlations respectively.

These definitions for $\SingleEventAvgDiffCorr{2}{n}$ and $\SingleEventAvgDiffCorr{4}{n}$ differ slightly from the standard $n^{\mathrm{th}}$-order $2$- and $4$-particle differential single event azimuthal correlations~\cite{Bilandzic:2010jr} due to the fact that we do not have to consider points which are labeled as both RFP and POI because the azimuthal angle of the jet axis $\psi_{\text{jet}}$ is a composite angle from recombination of particles using a jet clustering algorithm.  In practice, for jet $v_n$, it may be necessary to use RFP and POI correlations in different rapidity regions in order to avoid autocorrelations from using jet constituents in the RFP.

From the definitions for $\SingleEventAvgDiffCorr{2}{n}$ and $\SingleEventAvgDiffCorr{4}{n}$, the second and fourth order POI cumulants can be defined as 
\begin{equation}
\begin{aligned}
    d_{n}\{2\} \equiv \EventAvgDiffCorr{2}{n},
\end{aligned}
\label{Eq:dn2_def}
\end{equation}
and
\begin{equation}
\begin{aligned}
    d_{n}\{4\} \equiv \EventAvgDiffCorr{4}{n} - 2\cdot \EventAvgDiffCorr{2}{n}\EventAvgCorr{2}{n},
\end{aligned}
\label{Eq:dn4_def}
\end{equation}
where $d_{n}\{2\}$ is the the  $n^{\mathrm{th}}$-order $2$-particle average differential azimuthal correlation and $d_{n}\{4\}$ is the $n^{\mathrm{th}}$-order $4$-particle differential azimuthal correlation corrected for event-by-event fluctuations.

The reference azimuthal anisotropies are calculated from the 
reference cumulants $c_{n}\{2k\}$ defined in Equations~\ref{Eq:cn2_def}~\ref{Eq:cn4_def}
\begin{equation}
\begin{aligned}
    v_{n}\{2\} \equiv \sqrt{c_{n}\{2\} },
\end{aligned}
\label{Eq:vn2_def}
\end{equation}
and
\begin{equation}
\begin{aligned}
    v_{n}\{4\} \equiv \sqrt[4]{ -c_{n}\{4\} },
\end{aligned}
\label{Eq:vn4_def}
\end{equation}
where $v_{n}\{2\}$ and $v_{n}\{4\}$ are used to denote estimations of the integrated flow $v_{n}$ using the $2^{\text{nd}}$ and $4^{\text{th}}$-order cumulants respectively. 

The differential jet azimuthal anisotropies $v^{\text{jet}}_{n}\{2k\}$ are calculated using the differential cumulants  $d_{n}\{2k\}$ defined in Equations~\ref{Eq:dn2_def}~\ref{Eq:dn4_def} with respect to the the reference azimuthal anisotropies $v_{n}\{2k\}$ 
\begin{equation}
\begin{aligned}
    v^{\text{jet}}_{n}\{2\}(p_{T}) = \frac{d_{n}\{2\}(p_{T})}{v_{n}\{2\}},
\end{aligned}
\label{Eq:vn2_diff_def}
\end{equation}

\begin{equation}
\begin{aligned}
    v^{\text{jet}}_{n}\{2\}(p_{T}) = -\frac{d_{n}\{4\}(p_{T})}{(v_{n}\{4\})^{3/4}},
\end{aligned}
\label{Eq:vn4_diff_def}
\end{equation}
where $v^{\text{jet}}_{n}\{2\}(p_{T})$ and  $v^{\text{jet}}_{n}\{4\}(p_{T})$ are used to denote estimations of the differential flow for jets with a given transverse momentum $p_{T}$ using the $2^{\text{nd}}$ and $4^{\text{th}}$-order differential cumulants respectively.

\bibliography{main_refs}

\end{document}